\documentclass[preprint,showpacs,preprintnumbers,amsmath,amssymb]{revtex4}

\usepackage{graphicx}
\usepackage{dcolumn}
\usepackage{bm}
\usepackage{hyperref}

\newcommand{\qed}{\nobreak \ifvmode \relax \else
      \ifdim\lastskip<1.5em \hskip-\lastskip
      \hskip1.5em plus0em minus0.5em \fi \nobreak
      \vrule height0.75em width0.5em depth0.25em\fi}

\begin{document}

\preprint{}

\title{Toward Exact $2 \times 2$ Hilbert-Schmidt Determinantal Probability Distributions {\it via} Mellin Transforms and Other Approaches}
\author{Paul B. Slater}%
\email{slater@kitp.ucsb.edu}
\affiliation{%
University of California, Santa Barbara, CA 93106-4030\\
}%
\date{\today}

\begin{abstract}
We attempt to construct the exact univariate probability distributions for $2 \times 2$ quantum systems that yield
the (balanced) univariate Hilbert-Schmidt determinantal moments 
$\left\langle (\left\vert \rho\right\vert \left\vert \rho^{PT}\right\vert)
^{n}\right\rangle $, obtained by Slater and Dunkl ({\it J. Phys. A}, {\bf{45}}, 095305 [2012]). To begin, we follow--to the extent possible--the Mellin transform-based approach of Penson and {\.Z}yczkowski in their study of Fuss-Catalan and Raney distributions  ({\it Phys. Rev. E}, {\bf{83}}, 061118 [2011]). Further,  we approximate the $y$-intercepts (separability/entanglement boundaries)--at which 
$|\rho^{PT}|=0$-- of the probability distributions based on the (balanced) moments, as well as the previously reported unbalanced determinantal moments $\left\langle \left\vert \rho^{PT}\right\vert
^{n}\right\rangle $,  as a function of the seventy values of the Dyson-index-like parameter 
$\alpha = \frac{1}{2}$ (rebits), 1 (qubits), $\frac{3}{2}, 2$ (quaterbits) $\ldots,35$.

\end{abstract}


\maketitle
Slater and Dunkl reported formulas--involving (generalized) 
hypergeometric functions 
($_{p}F_{p-1}$)--for the Hilbert-Schmidt moments of two sets of univariate probability distributions pertaining to the entanglement/separability of 
$2 \times 2$ quantum systems \cite[p. 30]{MomentBased} (cf. \cite{BuresHilbert}). One ("balanced") set of (determinantal) moments  had the form ($\rho$ denoting a $4 \times 4$ density matrix, and $\rho^{PT}$, its partial transpose) 
\begin{gather*} \label{nequalk}
\left\langle (\left\vert \rho\right\vert \left\vert \rho^{PT}\right\vert)
^{n}\right\rangle \\
=\frac{\left(  2n\right)  !\left(  1+\alpha\right)  _{2n}\left(
1+2\alpha\right)  _{2n}}{2^{12n}\left(  3\alpha+\frac{3}{2}\right)
_{2n}\left(  6\alpha+\frac{5}{2}\right)  _{4n}}~_{4}F_{3}\left(
\genfrac{}{}{0pt}{}{\ -n,\alpha,\alpha+\frac{1}{2},-4n-1-5\alpha
}{-2n-\alpha,-2n-2\alpha,\frac{1}{2}-n}%
;1\right)  .
\end{gather*}
and the other set of ("unbalanced") determinantal moments, the form
\begin{gather*} \label{nequalzero}
\left\langle \left\vert \rho^{PT}\right\vert ^{n}\right\rangle =\frac
{n!\left(  \alpha+1\right)  _{n}\left(  2\alpha+1\right)  _{n}}{2^{6n}\left(
3\alpha+\frac{3}{2}\right)  _{n}\left(  6\alpha+\frac{5}{2}\right)  _{2n}}\\
+\frac{\left(  -2n-1-5\alpha\right)  _{n}\left(  \alpha\right)  _{n}\left(
\alpha+\frac{1}{2}\right)  _{n}}{2^{4n}\left(  3\alpha+\frac{3}{2}\right)
_{n}\left(  6\alpha+\frac{5}{2}\right)  _{2n}}~_{5}F_{4}\left(
\genfrac{}{}{0pt}{}{-\frac{n-2}{2},-\frac{n-1}{2},-n,\alpha+1,2\alpha
+1}{1-n,n+2+5\alpha,1-n-\alpha,\frac{1}{2}-n-\alpha}%
;1\right)  .
\end{gather*}
(Because of the denominator parameter $1-n$ it is necessary to replace the
$_{5}F_{4}$-sum by $1$ to obtain the correct value when $n=1$. Both hypergeometric functions are terminating (as well as  {\it balanced} in the Pfaff-Saalschutzian sense \cite[sec. 2.2]{bailey} in nature.)
Here $\alpha$ is a Dyson-index-like parameter that takes the value $\frac{1}{2}$, 1 and 2 for real, complex and quaternionic systems, respectively.

The probability distributions having the first set of balanced moments extend over the range $[-2^{-12} \cdot  3^{-3}, 2^{-16}]$, with the nonnegative interval  
$[0,2^{-16}]$ corresponding to separable systems.
The probability distributions possessing the second set of unbalanced moments extend over $[-2^{-4},2^{-8}]$, with the nonnegative interval $[0,2^{-8}]$ corresponding to separable systems.

We are interested in finding the exact probability distributions yielding these sets of moments--the accomplishment of which task would presumably yield further insight into the nature and character of the derived 
separability probabilities. In undertaking such an effort, to begin, we follow--as best we can--the analytical framework, involving Mellin transforms and Meijer $G$ functions, employed by Penson and {\.Z}yczkowski in their study of Fuss-Catalan and Raney distributions  \cite{twoKarols}. 

We have largely worked with the second set of (unbalanced) moments in recent efforts of ours \cite{SlaterHyper,SlaterConcise}. In fact, in \cite{SlaterConcise}, making use of this particular set, we were able to report the following "concise" formula for the probability $P(\alpha)$ that a $2 \times 2$ system is separable (that is, the cumulative probability
over [either of] the indicated nonnegative intervals)
\begin{equation} \label{Hou1}
P(\alpha) =\Sigma_{i=0}^\infty f(\alpha+i),
\end{equation}
where
\begin{equation} \label{Hou2}
f(\alpha) = P(\alpha)-P(\alpha +1) = \frac{ q(\alpha) 2^{-4 \alpha -6} \Gamma{(3 \alpha +\frac{5}{2})} \Gamma{(5 \alpha +2})}{3 \Gamma{(\alpha +1)} \Gamma{(2 \alpha +3)} 
\Gamma{(5 \alpha +\frac{13}{2})}},
\end{equation}
and
\begin{equation} \label{Hou3}
q(\alpha) = 185000 \alpha ^5+779750 \alpha ^4+1289125 \alpha ^3+1042015 \alpha ^2+410694 \alpha +63000 = 
\end{equation}
\begin{displaymath}
\alpha  (5 \alpha  (25 \alpha  (2 \alpha  (740 \alpha
   +3119)+10313)+208403)+410694)+63000.
\end{displaymath}
It was, then, concluded (using strongly convincing numerical evidence consisting of expansions of hundreds of digits) that for the (nine-dimensional) two-rebit systems, $P(\frac{1}{2})= \frac{29}{64}$, for the (fifteen-dimensional) two-qubit systems, $P(1) =\frac{8}{33}$, and for the (twenty-seven-dimensional) two-quater(nionic)-bit systems, $P(2)= \frac{26}{323}$.

Nevertheless, despite these substantial successes with the particular use of the second set (convergence of probability-distribution reconstruction procedures \cite{Provost} being much slower with the first set), it appears that the first set of  moments is more immediately amenable to the implementation of the Penson-{\.Z}yzckowski approach for reconstruction of the complete probability distributions of interest.

\begin{figure}
\includegraphics{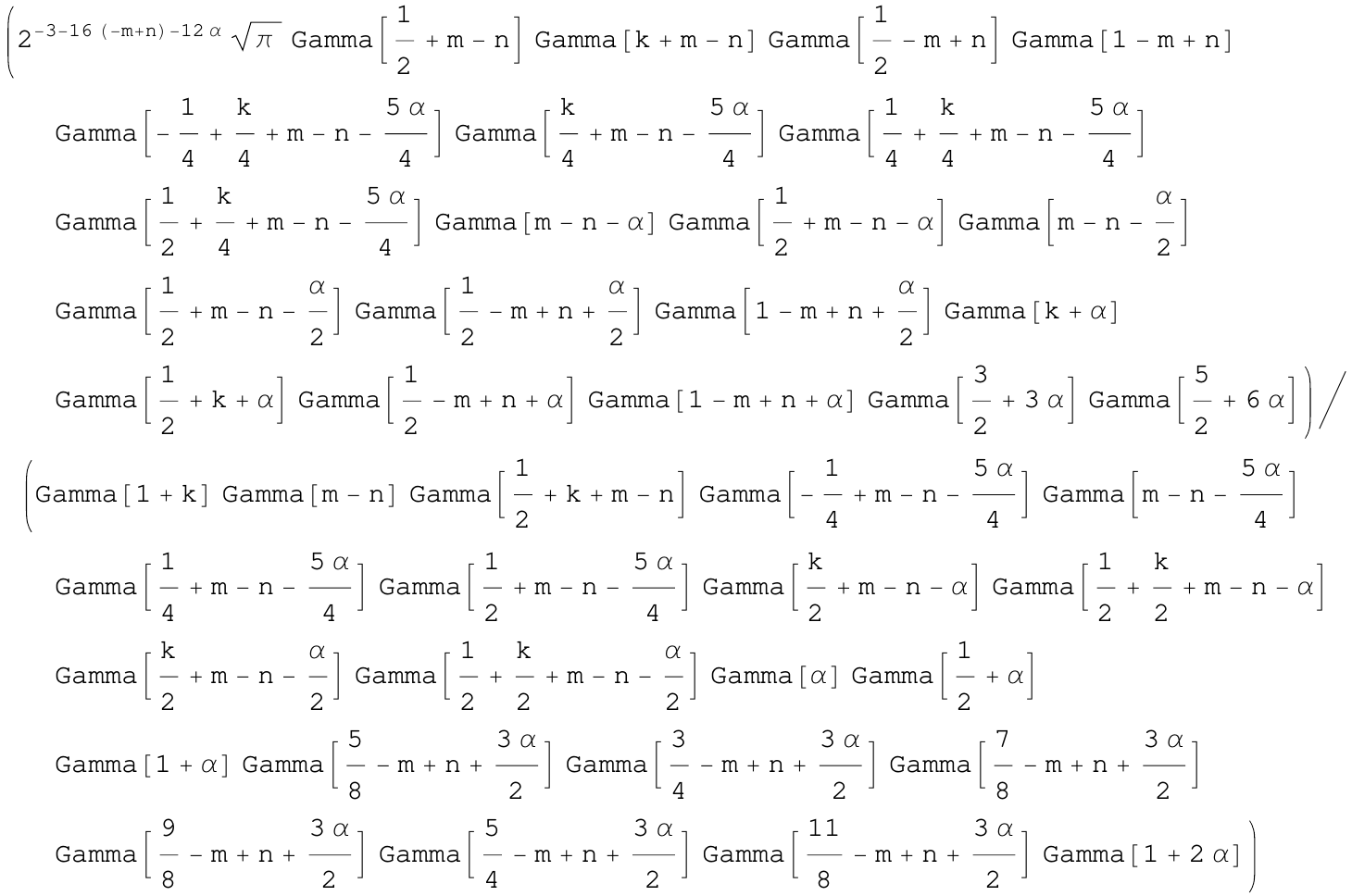}
\caption{\label{fig:GaussMultiplication}Quantity which when summed over $k$ from 0 to $n-m$ yields $\left\langle (\left\vert \rho\right\vert \left\vert \rho^{PT}\right\vert)^{n-m}\right\rangle $}
\end{figure}

\begin{figure}
\includegraphics{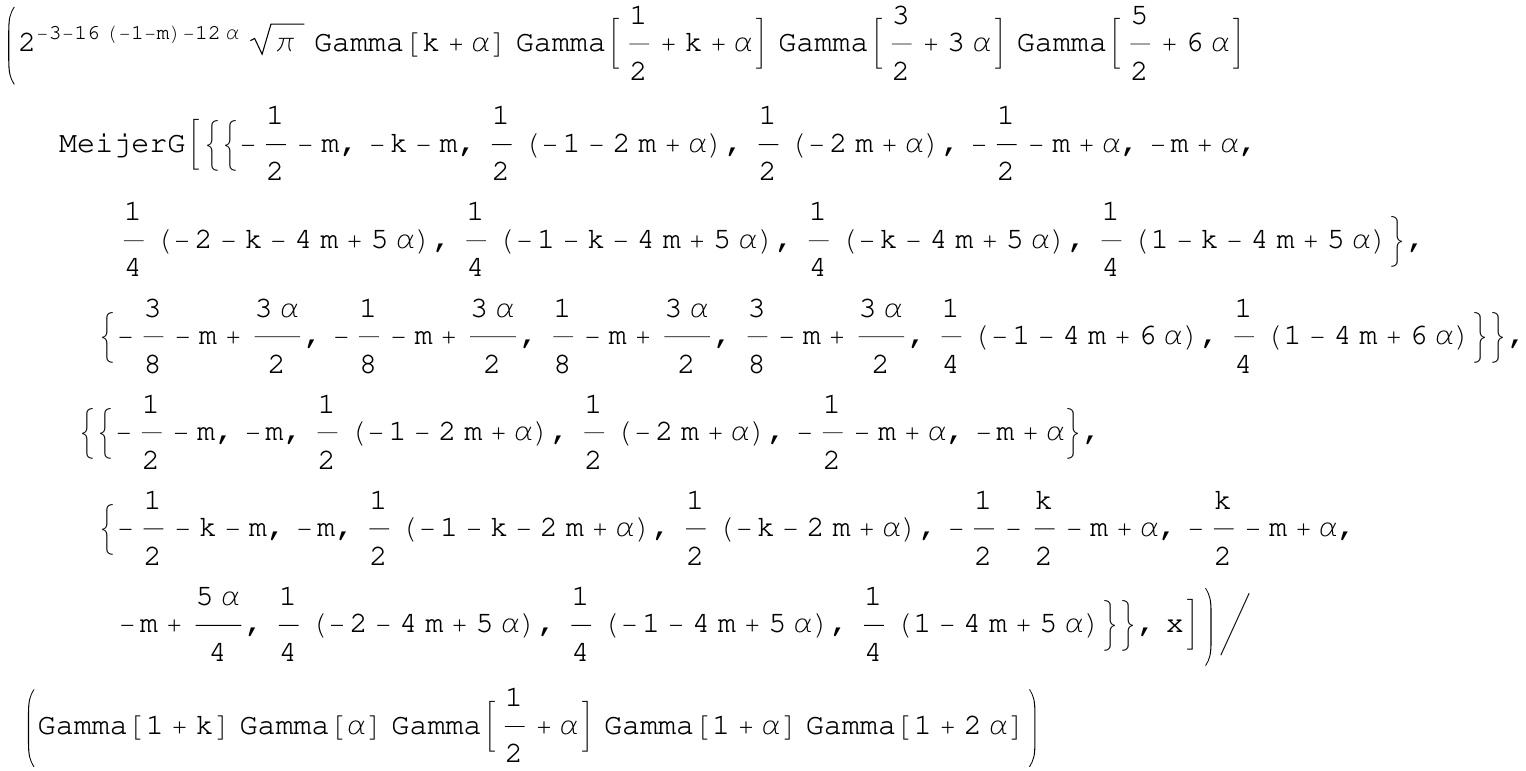}
\caption{\label{fig:MeijerTerm}Inverse Mellin transform of quantity in Fig.~\ref{fig:GaussMultiplication}.}
\end{figure}

Since (quite significantly, it would appear) the ranges of the indicated probability distributions include negative-valued intervals, it appears necessary--for application of the Mellin transform--to appropriately modify the distributions to eliminate nonnegative domains. If we do linearly transform the balanced moments of the probability distributions defined over $[-2^{-12} \cdot 3^{-3}, 2^{-16}]$ to the moments of probability distributions defined over the (nonnegative) unit interval [0,1], we obtain for the new $n$-th moments (using the binomial expansion)
\begin{equation}
\left\langle (\left\vert \rho\right\vert \left\vert \rho^{PT}\right\vert)
^{n}\right\rangle_{[0,1]}= 
\Big(\frac{1}{2^{-12} \cdot 3^{-3} + 2^{-16}}\Big)^n \Sigma_{m=0}^n {n \choose m} (2^{12} \cdot 3^{3})^{-m} \left\langle (\left\vert \rho\right\vert \left\vert \rho^{PT}\right\vert)
^{n-m}\right\rangle.
\end{equation} 
(Of course, it would be highly desirable to have a more explicit expression for $\left\langle (\left\vert \rho\right\vert \left\vert \rho^{PT}\right\vert)
^{n}\right\rangle_{[0,1]}$ than this one--a task we are pursuing.) Now, (the original, untransformed) $\left\langle (\left\vert \rho\right\vert \left\vert \rho^{PT}\right\vert)
^{n-m}\right\rangle $  moment occurring in the right-hand side can be expressed--using the expansion formula for hypergeometric functions, as well as the Gauss multiplication theorem (for subsequent use of the inverse Mellin transform)--as the sum over $k$ from 0 to $n-m$ of the quantity in  Fig.~\ref{fig:GaussMultiplication}.
(Fig.~\ref{fig:DifferenceRootExpression} gives the function--setting $\alpha=1$--which when summed over $k$ from 0 to $n$, then added to $(2^{12} \cdot 3^3)^{-n}$, with the resultant sum  multiplied by ($\frac{1}{2^{-12} \cdot 3^{-3} +2^{-16}})^n$, yields the unit-interval moments 
$\left\langle (\left\vert \rho\right\vert \left\vert \rho^{PT}\right\vert)^n\right\rangle_{[0,1]} $.)
\includegraphics[page=1,scale=0.9]{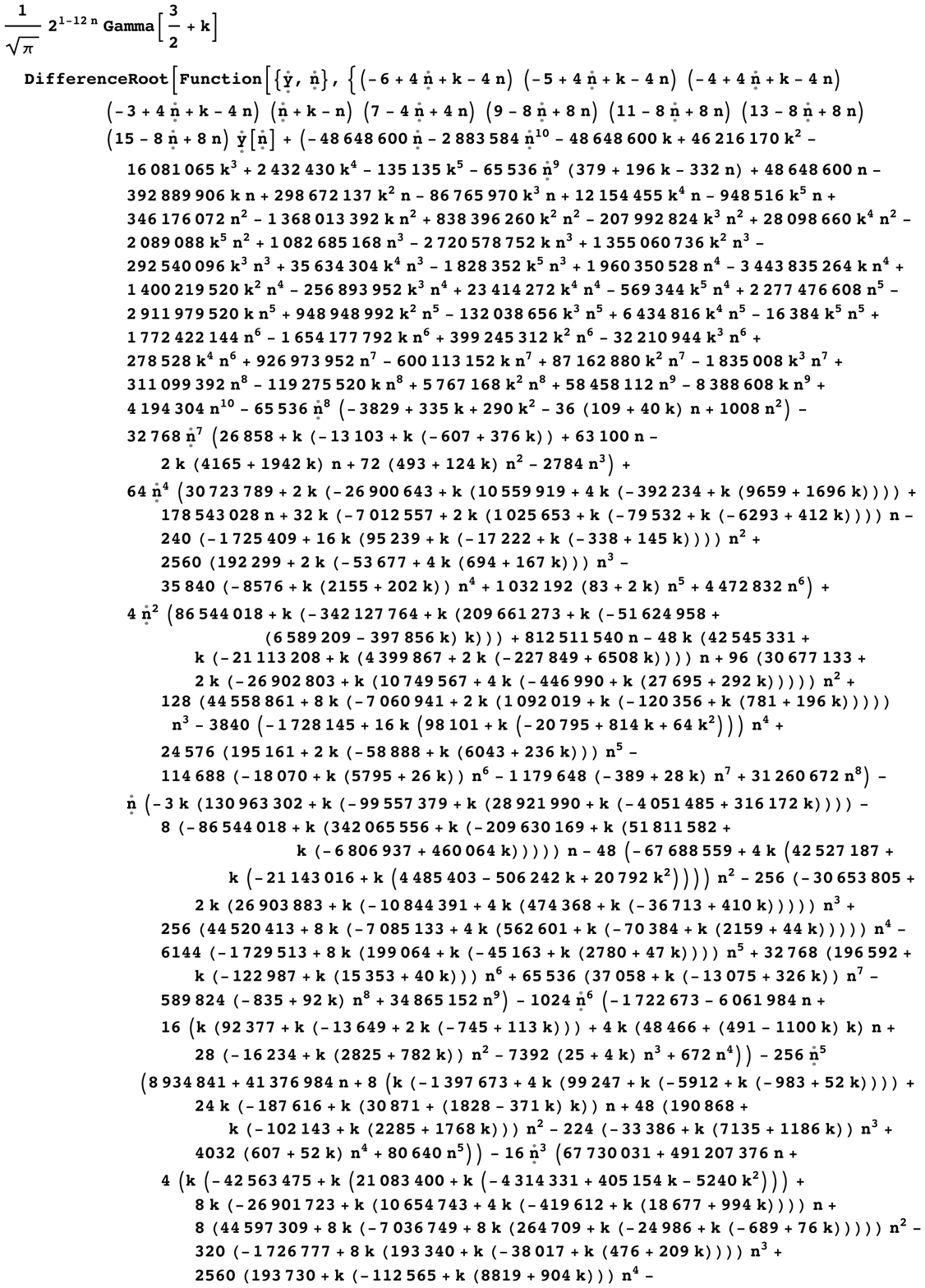}
\includegraphics[page=2,scale=0.9]{DifferenceRootExpression.pdf}
\begin{figure}
\caption{\label{fig:DifferenceRootExpression}The function--solving the indicated linear difference equation--which when summed over $k$ from 0 to $n$, then added to $(2^{12} \cdot 3^3)^{-n}$, with the resultant  sum  multiplied by ($\frac{1}{2^{-12} \cdot 3^{-3} +2^{-16}})^n$, yields the transformed (unit-interval) moments $\left\langle (\left\vert \rho\right\vert \left\vert \rho^{PT}\right\vert)^n\right\rangle_{[0,1]} $ in the two-qubit case $\alpha=1$.}
\end{figure}
Applying the inverse Mellin transform to the quantity in Fig.~\ref{fig:GaussMultiplication}, having made the required transformation $n \rightarrow \sigma-1$, we obtain the expression in 
Fig.~\ref{fig:MeijerTerm}.
Then, Fig.~\ref{fig:MeijerTermSimple} shows an equivalent form--in terms of hypergeometric functions--to the Meijer $G$ term in 
Fig.~\ref{fig:MeijerTerm} with the summation index $m$  having been set to zero (cf. \cite{twoKarols}).
\includegraphics[page=1,scale=0.9]{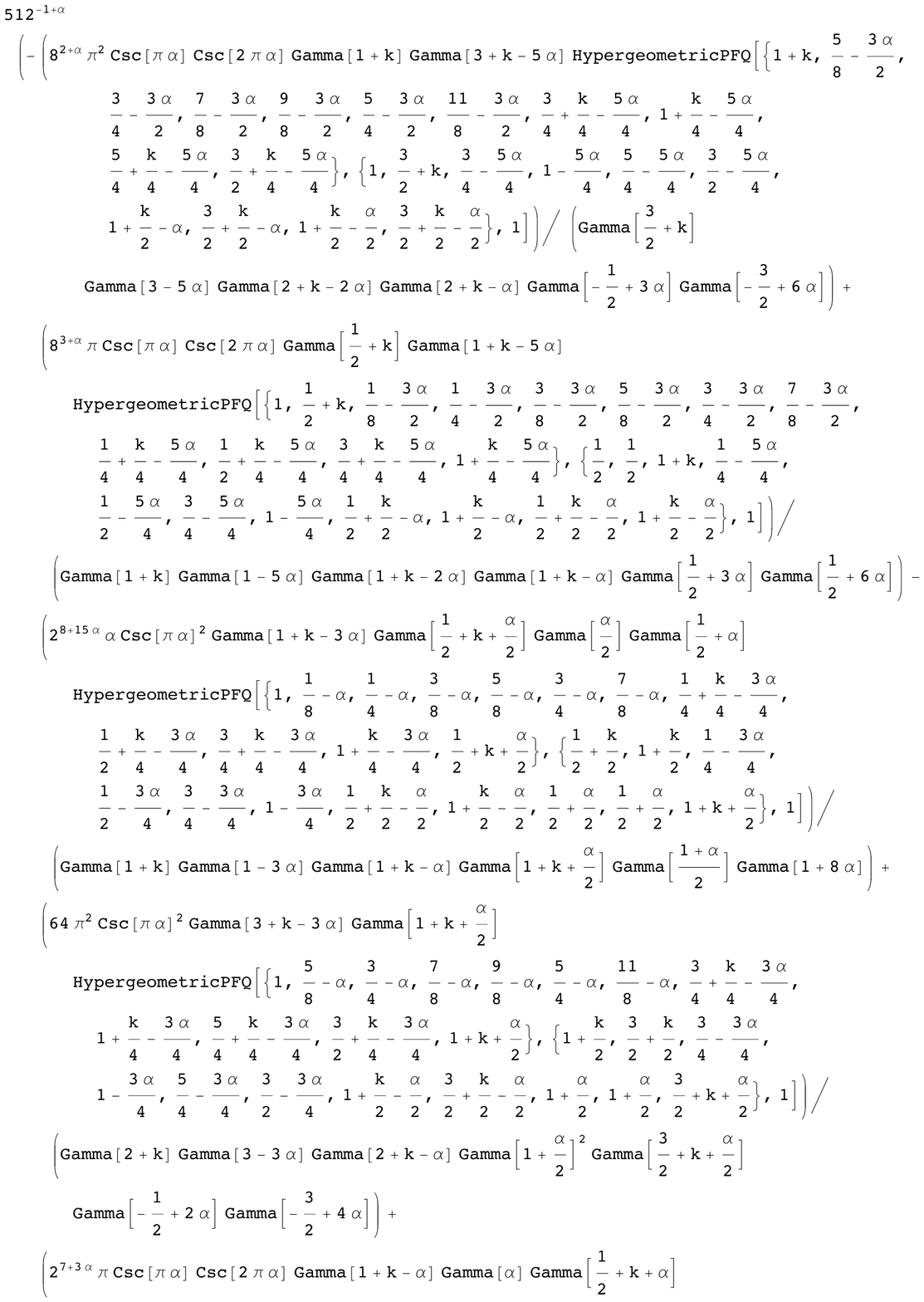}
\includegraphics[page=2,scale=0.9]{MeijerTermSimple.pdf}
\begin{figure}
\caption{\label{fig:MeijerTermSimple}Equivalent form--in terms of hypergeometric formulas--to the Meijer $G$ function term in 
Fig.~\ref{fig:MeijerTerm} with the summation index $m$ having been set to zero.}.
\end{figure}

So, certainly highly formidable obstacles remain to the achieving of our stated goal of explicitly constructing the exact univariate probability distributions--parameterized by the Dyson-index-like parameter $\alpha$--for $2 \times 2$ quantum systems that yield
the (balanced) univariate Hilbert-Schmidt determinantal moments 
$\left\langle (\left\vert \rho\right\vert \left\vert \rho^{PT}\right\vert)
^{n}\right\rangle $obtained by Slater and Dunkl \cite{MomentBased}.

One immediate issue to be addressed--so that Mellin transform methods might be more readily employed--is the development of  explicit formulas for the determinantal moments, transformed--to avoid negative domains--so that they correspond to probability distributions over the range $[0,1]$. Perhaps it is possible to exploit the fact that the hypergeometric functions in both sets of moments presented at the outset of the paper, are balanced (in the Pfaff-Saalschutzian sense \cite[sec. 2.2]{bailey} ) and terminating in character. Further, the use of Euler's integral representation of the generalized hypergeometric function \cite{andrews} might prove of value.

\begin{figure}
\includegraphics{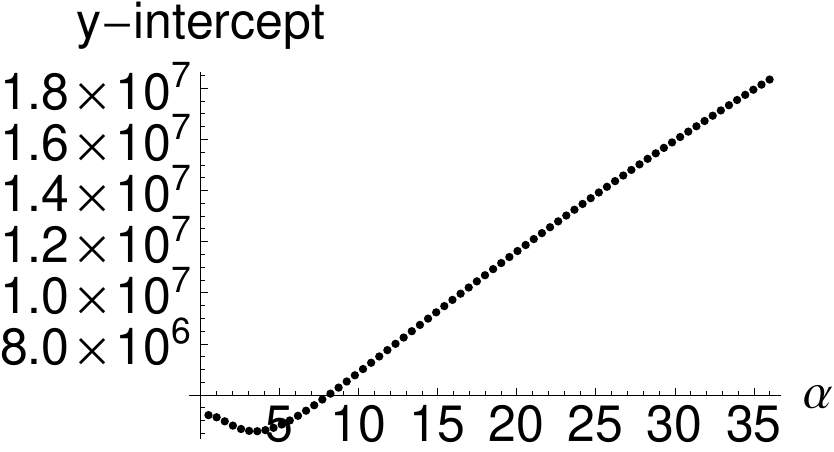}
\caption{\label{fig:InterceptBalanced}The $y$-intercepts--at which 
$|\rho^{PT}|=0$--as a function of the seventy values of the Dyson-index-like parameter $\alpha = \frac{1}{2}, 1, \frac{3}{2},\ldots,35$, for the class of probability distributions, extending over  
$[-2^{-12} \cdot  3^{-3}, 2^{-16}]$ based on the first (balanced) set of moments.  The first 1900 determinantal moments were employed in the probability-distribution Legendre-polynomial-based reconstruction process}.
\end{figure}

\begin{figure}
\includegraphics{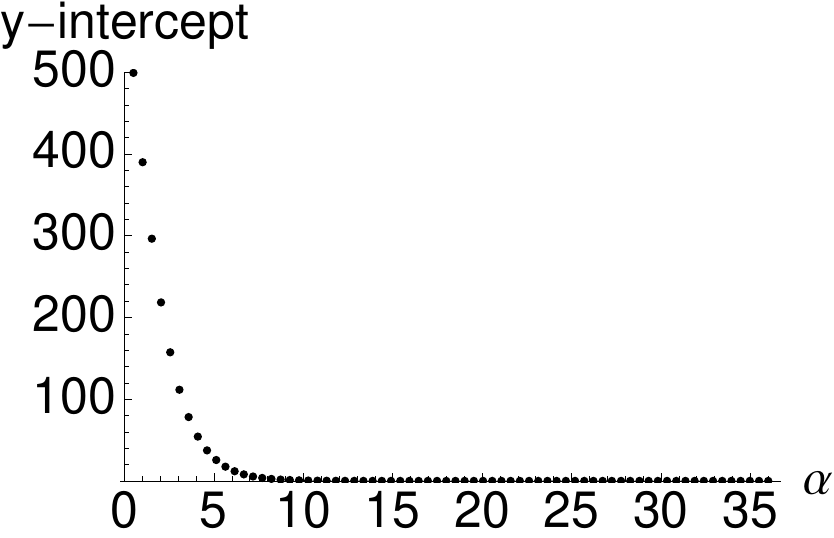}
\caption{\label{fig:Intercept}The $y$-intercepts--at which $|\rho^{PT}|=0$--as a function of the seventy values of the Dyson-index-like parameter $\alpha = \frac{1}{2}, 1, \frac{3}{2},\ldots,35$, for the class of probability distributions, extending over  
$[-2^{-4},2^{-8}]$ based on the second (unbalanced) set of moments.  The first 2000 determinantal moments were employed in the probability-distribution Legendre-polynomial-based reconstruction process}.
\end{figure}

As another approach to ascertaining the properties of the probability distribution functions in question, we have used the Mathematica-implemented Legendre-polynomial-based reconstruction algorithm of Provost \cite{Provost}, that we have previously applied with considerable success \cite{MomentBased,SlaterHyper}, to determining the $y$-intercepts of the Hilbert-Schmidt determinantal probability density functions. That is, we seek the values of these probability density functions at which (the $x$-variable) 
$|\rho^{PT}|$ is zero.
In Fig.~\ref{fig:InterceptBalanced} we show the seventy $y$-intercepts, as a function of the Dyson-index-like parameter $\alpha = \frac{1}{2}$ (rebits), 1 (qubits), $\frac{3}{2},2$ (quaterbits), $\ldots,35$, for the class of probability distributions, extending over  
$[-2^{-12} \cdot  3^{-3}, 2^{-16}]$ based on the first (balanced) set of moments.  (The first 1900 determinantal moments were employed.)
In Fig.~\ref{fig:Intercept} we show the seventy $y$-intercepts, as a function of the Dyson-index-like parameter $\alpha = \frac{1}{2}, 1, \frac{3}{2},\ldots,35$, for the class of probability distributions, extending over  
$[-2^{-4},2^{-8}]$ based on the second (unbalanced) set of moments.  (The first 2000 determinantal moments were employed.)
We hope to be able to discern (increasing further the numbers of moments employed) exact values for these $y$-intercepts, which would then hopefully cast light on the specific nature of the Hilbert-Schmidt determinantal probability density functions that have been the subject of this communication.

\begin{figure}
\includegraphics{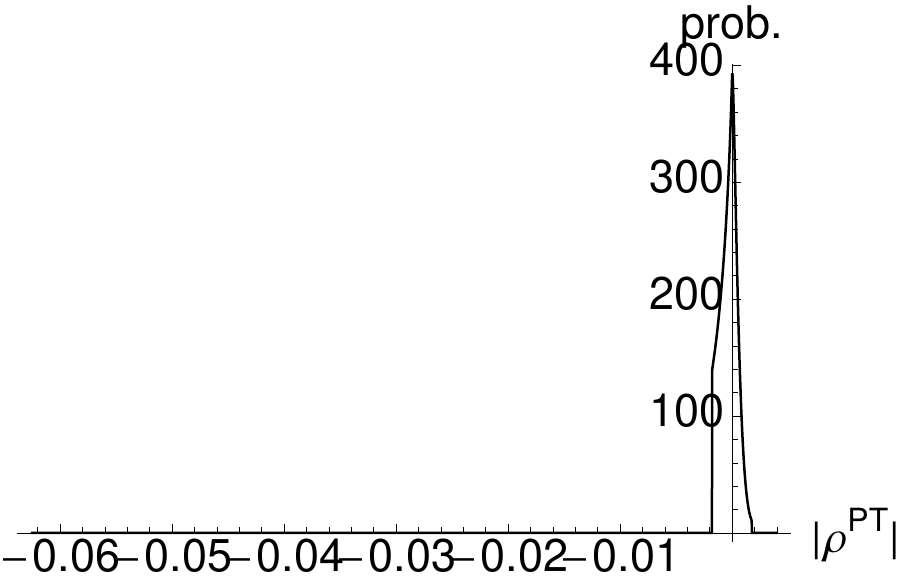}
\caption{\label{fig:FullCurve}Approximation--based on the first 1250 unbalanced moments, having set $\alpha=1$--to the two-qubit separability probability distribution over $[-\frac{1}{16},\frac{1}{256}]$.}
\end{figure}

\begin{figure}[H]
\includegraphics{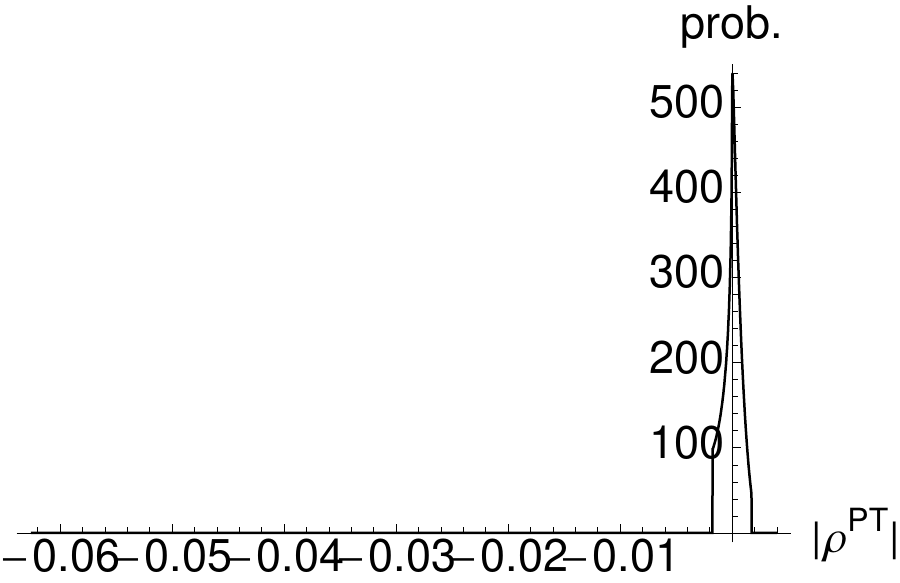}
\caption{\label{fig:FullCurveRebit}Approximation--based on the first 1250 unbalanced moments, having set $\alpha=\frac{1}{2}$--to the two-rebit separability probability distribution over $[-\frac{1}{16},\frac{1}{256}]$.}
\end{figure}

\begin{figure}[H]
\includegraphics{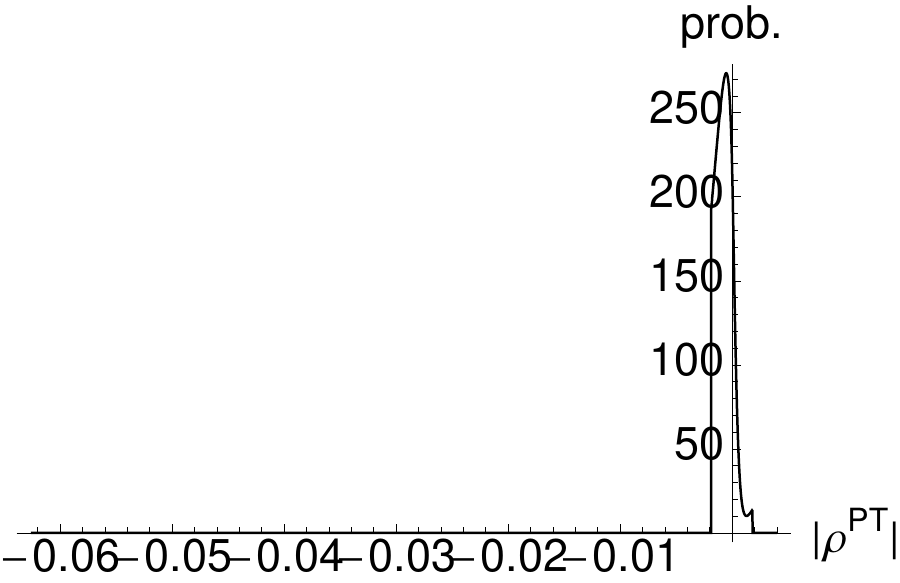}
\caption{\label{fig:FullCurveQuaterbit}Approximation--based on the first 1250 unbalanced moments, having set $\alpha=2$--to the two-quaterbit separability probability distribution over $[-\frac{1}{16},\frac{1}{256}]$.}
\end{figure}

Another possible use of the Legendre-polynomial-based probability-density reconstruction process \cite{Provost} might be to determine the modes of the yet-unknown separability probability density functions as functions of 
$\alpha$. Along such lines, in Fig.~\ref{fig:FullCurve}, we present--based on the first 1250 unbalanced moments, setting $\alpha=1$, an approximation to the two-qubit probability distribution defined over
$[-\frac{1}{16},\frac{1}{256}]$. (An estimate--based upon the first 500 moments--of the median of the distribution is $|\rho^{PT}|=-0.00691863$.) The cumulative (separability) probability over the nonnegative interval $[0, \frac{1}{256}]$ appears to be equal to $\frac{8}{33}$, as we have recently argued \cite{MomentBased,SlaterHyper,SlaterConcise} (and indicated at the outset of this paper).
In Fig.~\ref{fig:FullCurveRebit}, we show the two-rebit counterpart ($\alpha = \frac{1}{2}$) to this  curve, and in Fig.~\ref{fig:FullCurveQuaterbit} the two-quaterbit counterpart ($\alpha=2$). (Estimates of the medians--based upon the first 500 moments--of the last two distributions are, respectively, 
$|\rho^{PT}|=-0.00562687$ and $|\rho^{PT}|=-0.0121435$.)

\begin{figure}[H]
\includegraphics[scale=.8]{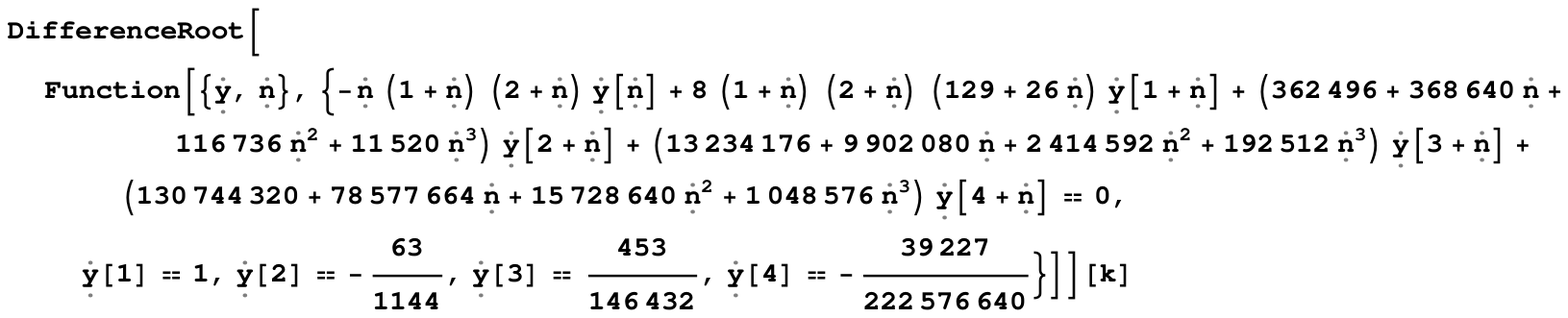}
\caption{\label{fig:TransformedMoments}Formula satisfied by the two-rebit ($\alpha=\frac{1}{2}$) Hilbert-Schmidt moments after their transformation so that the original range $[0,\frac{1}{256}]$ of the determinant of the density matrix matches the range $[-\frac{1}{16},\frac{1}{256}]$ of the determinant of the partial transpose.}
\end{figure}

For the two-rebit ($\alpha=\frac{1}{2}$) systems, we have 
the (univariate Hilbert-Schmidt determinantal moment) formulas \cite{csz}[eq. (3.2)] \cite[eq. (1)]{MomentBased}
(cf. \cite[Theorem 4]{andai}):
\begin{equation} \label{firstold}
\left\langle |\rho|^k \right\rangle =945 \Big( 4^{3-2 k} 
\frac{ \Gamma (2 k+2) \Gamma (2 k+4)}{\Gamma (4 k+10)} \Big).
\end{equation}
Now, $|\rho| \in [0, \frac{1}{256}]$. It might be analytically convenient 
(for convolution purposes,\ldots [cf. \cite[App. B]{dunkl}]), at some point, to linearly transform this variable to possess the same range as $|\rho^{PT}|$, that is, 
$[-\frac{1}{16},\frac{1}{256}]$ (as opposed to transforming the variable $|\rho^{PT}|$  to range over the unit interval). Under such a transformation, the new forms of the moments (\ref{firstold}) are given by
the function in Fig.~\ref{fig:TransformedMoments}, solving the indicated linear difference equation. 
Further, the probability distribution function ($f(y)$, with $y= 17 |\rho|-
\frac{1}{16}$)
yielding these moments is (cf. \cite[App. D.2]{MomentBased})
\begin{equation}
f(y)=-\frac{4128768 \sqrt{17-4 \sqrt{272 y+17}} y}{289
   \sqrt{17}}-\frac{72576}{289} \sqrt{16 y+1} \sqrt{17-4 \sqrt{272
   y+17}}
\end{equation}
\begin{displaymath}
-\frac{189504 \sqrt{17-4 \sqrt{272 y+17}}}{289 \sqrt{17}}
\end{displaymath}
\begin{displaymath}
+\frac{7741440}{289} y \tanh ^{-1}\left(\sqrt{1-\frac{4 \sqrt{16
   y+1}}{\sqrt{17}}}\right)+\frac{483840}{289} \tanh
   ^{-1}\left(\sqrt{1-\frac{4 \sqrt{16 y+1}}{\sqrt{17}}}\right).
\end{displaymath}

\begin{acknowledgments}
I would like to express appreciation to the Kavli Institute for Theoretical
Physics (KITP)
for computational support in this research. 
\end{acknowledgments}

\bibliography{Meijer3}

\end{document}